\documentclass[]{aa}
\def\etal{{et\,al. }}
\usepackage{graphicx}
\begin{document}
   \title{INTEGRAL Spectrometer SPI's GRB detection capabilities\thanks{Based on observations with INTEGRAL, an ESA project
with instruments and science data centre funded by ESA
member states (especially the PI countries: Denmark, France,
Germany, Italy, Switzerland, Spain), Czech Republic and
Poland, and with the participation of Russia and the USA.}
}

   \subtitle{GRBs detected inside SPI's FoV and with the anticoincidence system ACS}

   \author{
 A. von Kienlin\inst{1} \and
 V. Beckmann\inst{3,4} \and
 A. Rau \inst{1}   \and
 N. Arend \inst{1} \fnmsep\thanks{\emph{Present address:}
TU M\"unchen, Department of Physics, Institute E21,
James-Franck-Str., 85748 Garching, Germany} \and
 K. Bennett \inst{9} \and
 B. McBreen \inst{5} \and
 P. Connell \inst{6} \and \\
 S. Deluit \inst{4} \and
 L. Hanlon \inst{5} \and
 M. Kippen \inst{8} \and
 G. G. Lichti \inst{1} \and
 L. Moran \inst{5} \and
 R. Preece \inst{7} \and
 J.-P. Roques \inst{2} \and 
 V. Sch\"{o}nfelder \inst{1} \and
 G. Skinner \inst{2} \and
 A. Strong \inst{1} \and
 R. Williams \inst{9} 
}

   \offprints{A. von Kienlin \\ (\email{azk@mpe.mpg.de}) }

   \institute{ 
Max-Planck-Institut f\"{u}r extraterrestrische Physik, Giessenbachstrasse, 85748 Garching, Germany, \and
Centre d'Etudes Spatiales des Rayonnements, 9, avenue du Colonel Roche, 31028 Toulouse, France, \and
Institut f\"ur Astronomie und Astrophysik, Universit\"at T\"ubingen, Sand 1, D-72076 T\"ubingen, Germany \and
INTEGRAL Science Data Centre, Chemin d' \'Ecogia 16, CH-1290 Versoix, Switzerland \and
Department of Experimental Physics, University College Dublin, Stillorgan Road, Dublin 4, Ireland \and
Space Research Group, Univ. of Birmingham B15 2TT, Birmingham, United Kingdom \and
Dept. of Physics, UAH, National Space Science \& Technology Center, Huntsville, AL  35805, USA\and
Space and Remote Sensing Sciences, NIS-2 Mail Stop D436, Los Alamos National Laboratory, Los Alamos, NM 87545, USA\and
Science Operations and Data Systems Division of ESA/ESTEC, SCI-SDG, Keplerlaan 1, Postbox 299, NL-2200 AG Noordwijk, The Netherlands}

   \date{Received xxx 00, 2003; accepted xxx 00, 2003}

   \abstract{
The spectrometer SPI, one of the two main instruments of the INTEGRAL spacecraft,
offers significant  gamma-ray burst detection capabilities. In its 35$^o$ (full width) 
field of view  SPI is able to localise gamma-ray bursts at a mean rate of $\sim$\,0.8/month. With its large 
anticoincidence shield  of 512 kg of BGO crystals SPI is able to detect gamma-ray bursts quasi omni-directionally 
with a very high sensitivity. Burst alerts of the anticoincidence shield are distributed by the INTEGRAL Burst Alert System. 
In the first 8 months of the mission about 0.8/day gamma-ray burst candidates and 0.3/day gamma-ray burst positions were obtained 
with the anticoincidence shield  by interplanetary network triangulations with other spacecrafts.

   }

   \maketitle
%

\section{Introduction}   

\label{sect:intro}  

The detection and investigation of cosmic gamma-ray bursts (GRBs) is an important scientific topic of the INTEGRAL
mission. Even now, 35 years after 
their discovery by the Vela satellites (\cite{Klebesadel73}), this phenomenon is far from being well-understood.
A major breakthrough in this field was obtained in the 1990's with the BATSE detectors (\cite{Fishman89}) 
on NASA's Compton Gamma-Ray Observatory (CGRO). In the ten years of the CGRO mission 2704 bursts were
registered, which showed an isotropic distribution over the entire sky, but with a deficiency for weak burst 
compared to a homogeneously distribution (logN-logS distribution with $-3/2$ power law).
It was the Italian/Dutch satellite BeppoSAX (\cite{Boella97}) which helped to reveal the cosmological nature of GRBs with
the identification of the first X-ray afterglow (\cite{Costa97}), which triggered the first successful follow-up
observation at optical wavelengths (\cite{Paradijs97})  and subsequent direct redshift measurements 
(\cite{metzger97}). The redshifts obtained to date for about 30 GRBs range from 0.0085 to
4.5. Observational evidence now strongly suggests that GRBs 
longer than 2 seconds are associated with hypernovae (\cite{galama98,kawabata03,hjorth03}). 
Gamma-ray polarimetry of the intense event GRB021206 
using the RHESSI satellite (\cite{CobBoggs2003}) indicates that the GRB emission is highly
polarised (80\%), suggestive of synchrotron radiation from particles in a highly-ordered magnetic field (\cite{granot03}). 

\section{GRB detection with the camera of SPI}

The aim of the spectrometer SPI  is to perform high-resolution
spectroscopy of astrophysical sources in the energy range between
20\,keV and 8\,MeV (\cite{spi}). The imaging capability is good,
but is exceeded by that of the imager IBIS  which complements SPI
by having higher imaging resolution, but lower spectroscopic
resolving power (\cite{ibis}). Hence we expect the two main
INTEGRAL instruments to make contributions to GRB research in
different areas. The connection between GRBs and Type Ib/c
`hypernovae' (\cite{hjorth03}) heralds a
potentially
exciting era for high-resolution $\gamma$-ray spectroscopy of the
closest of these sources. In addition, the broad energy coverage
of SPI is well suited to constrain the spectral shape, both below
and above the energy at which the GRB power output is typically
peaked ($\sim 250$\,keV). The long-standing controversy over the
existence of short-lived spectral features in GRB spectra can also
be addressed by SPI's superb spectroscopic capabilities.
On the other hand, IBIS will provide precise GRB locations to the
community, which is important for the observation of GRB
afterglows.
In addition, the capability to cross-calibrate both spectra and
images  between the two experiments is extremely important,
particularly in the case of such short-lived events as GRBs, which
cannot be re-observed. Currently GRBs which occur inside SPI's field of view (FoV)
are detected and analysed offline. The implementation of an
automatic GRB detection algorithm for SPI (\cite{spie2003}) into
the INTEGRAL burst-alert system (IBAS, \cite{ibas}) is
planned.

\subsection{Imaging with the spectrometer SPI}

The camera of SPI, which consists of 19 cooled high-purity germanium detectors, residing in a cryostat,
is shielded on the side walls and rear side by a large anticoincidence shield (ACS). The FoV  
of the camera is defined by the upper opening of the ACS. The imaging capability of the instrument is 
attained by a passive-coded mask mounted 1.7\,m above the camera. By observing  a series of nearby pointing 
directions around the source (``dithering'') the imaging capability of SPI is improved by reducing ambiguity effects.  
For the normal mode of operation two dithering patterns are applied, a square one with $5 \times 5$ pointings 
and a hexagonal one with 7 pointings. Each pointing lasts about 30\,min. The slew between pointings on a dither pattern 
lasts as 5-10 minutes. Due to their short duration, dithering will not improve the imaging of SPI for 
bursts. But during its short duration a burst will be in most cases the brightest $\gamma$-ray source for SPI 
in the whole FoV,  so imaging with SPI is still possible. The case when a burst occurred during a slew can be 
investigated for GRB030131. Despite the modest angular resolution of SPI, which is of the order of  2.5$^{\circ}$, 
it is possible to locate the direction of bursts down to a few arcminutes. 

\subsection{Analysis methods for GRBs} 
\label{grb-analysis}

The analysis of GRBs in the SPI FoV has been performed by using the instrument-specific 
software, which has been developed by the SPI instrument team in collaboration with the ISDC (\cite{beckmann02}).
In both, image and spectral extraction, the analysis methods are similar and are therefore described here 
together. SPI is a background-dominated instrument. The overall count rate of the detector plane when
observing an empty field is about 880\,counts/s, while e.g. the Crab shows a SPI 
count rate of 33\,counts/s. Therefore a careful background handling is essential. In the
case of short events as GRBs, the best way to remove the background is to take the observation which 
encloses the GRB (excluding the time when the GRB was detectable in SPI) as a background. By applying 
the same parameters as for the GRB analysis a detector spectrum should be extracted. This detector spectrum
does not only contain the background, but will also include sources in the FoV and therefore allows
an optimised data reduction. The data for the GRB should be extracted from a short time range around the peak,
in order to achieve a good signal-to-noise ratio for the result. Source location and spectral extraction are
performed applying the iterative-removal-of-sources method, which is implemented in the SPIROS analysis software
(\cite{spiros}). For source location no a-priori knowledge of the GRB position
is necessary, while in order
to extract a spectrum, the best known (e.g. ISGRI derived) source position should be applied. 

\subsection{Results obtained for the first bursts observed inside SPI's FoV} 

Since the start of the mission, six GRBs have been observed within SPI's FoV. Tab.\,\ref{tab:SPI-GRBs} summarises the important quantities
derived with SPI. In all cases the GRB was detected by  IBIS and for comparison the important parameters found 
from the IBIS data are also shown. 
Below for each of the GRBs a short summary is given, especially with the emphasis on the SPI performance:

\begin{table*} 
\centering
\caption[]{Table of GBRs detected inside the FoV of SPI. For comparison the
results of IBIS are listed too. The data sets are taken from \cite{grb021125} for
GRB021125, from \cite{grb021219} for GRB021219, from \cite{grb030131} for
GRB030131, from \cite{Mereghetti2003} for GRB030227, from \cite{grb030320-kienlin} for
GRB030320 and from \cite{beckmann03} for GRB030501.
The first column lists the date of the GRB,  the second the duration of the burst. 
Column {\it ``GRB Location''} quotes for SPI and IBIS-ISGRI (and IBIS-PICsIT in
one case) the centre (in R.A. and Dec.) and radius $r$ of the error circle
(90\% confidence value)
together with  the signal-to-noise (S/N) ratios, with which the GRB was
detected.
The column {\it ``GRB Loc. Offset''} lists the the
offset angles of centre of the SPI-error circle ({\it SPI}) to the one
obtained by IBIS and the offset angle of the GRB (IBIS localisation) 
with respect to the INTEGRAL pointing direction ({\it Point.}). 
The next column specifies the peak flux in the 20 -- 200 keV energy range. The GRB fluence is given for the same 
energy range. The column {\it ``Photon Index''} gives the result of a model fit of the GRB spectrum with a power law.
Several fields of the table are left blank, because the corresponding values are not reported in the
references mentioned above. The fluxes and fluences in Crab units were
determined in other energy ranges than the standard 20--200\,keV range used
for the table (see text). 
}
\label{tab:SPI-GRBs}
\begin{tiny}
\begin{tabular}{|l||l|l|l|l|c|c|c|c|c|l|c|} 
\hline
\rule[-1ex]{0pt}{3.5ex} &  &  &  \multicolumn{4}{c|}{GRB Location} &  \multicolumn{2}{c|}{GRB Loc. Offset}  &  Peak Flux & Fluence 
& Photon \\
\cline{4-11}
\rule[-1ex]{0pt}{3.5ex}  \raisebox{1.5ex}[-1.5ex]{GRB} &
\raisebox{1.5ex}[-1.5ex]{Durat.} & \raisebox{1.5ex}[-1.5ex]{Instr.} & R.A. (J2000) 
& Dec. (J2000) & $r$ & S/N  &  SPI & Point. & [$\frac{\rm photons}{\rm cm^2s}$] & [$\frac{\rm erg}{\rm cm^2}\times 10^{-6}$] & Index  \\
\hline
\hline
\rule[-1ex]{0pt}{3.5ex}   &   & SPI & $19^{\rm h} 07^{\rm m} $ & $06^{\circ}
25'$ & 21' & 15.4 & & 
&  $3.22 \pm 0.50$ & $3.93 \pm 0.27$ & $1.88 \pm 0.10$  \\
\cline{3-7} \cline{10-12}
\rule[-1ex]{0pt}{3.5ex} \raisebox{1.5ex}[-1.5ex]{030501}
&\raisebox{1.5ex}[-1.5ex]{$\sim$\,40\,s}& 
ISGRI & $19^{\rm h} 05^{\rm m} 30^{\rm s}$ & $06^{\circ} 18' 26"$ & 3' &
10.0 & \raisebox{1.5ex}[-1.5ex]{19'} 
& \raisebox{1.5ex}[-1.5ex]{13.25$^{\circ}$} & $2.7 \pm 1.0$   & $3 \pm 1$ &  $1.75 \pm 0.10$ \\
\hline
\hline
\rule[-1ex]{0pt}{4ex}   &   & SPI & $17^{\rm h} 53.4^{\rm m} $ &
$-26^{\circ} 2.8'$ & 33' & 7.2 & & 
&  3.1  & $13.5{+2.1\atop -2.6}$ & $1.51 \pm 0.16$  \\
\cline{3-7} \cline{10-12}
\rule[-1ex]{0pt}{4ex} \raisebox{1.5ex}[-1.5ex]{030320} & \raisebox{1.5ex}[-1.5ex]{$\sim$\,60\,s} 
& ISGRI & $17^{\rm h} 51^{\rm m} 42^{\rm s}$ & $-25^{\circ} 18' 44"$ & 3'&15 & \raisebox{1.5ex}[-1.5ex]{49.8'} 
& \raisebox{1.5ex}[-1.5ex]{15.5$^{\circ}$} & 5.7   & 11 &  $1.69 {+0.07\atop -0.08} $ \\
\hline
\hline
\rule[-1ex]{0pt}{3.5ex}   &   & SPI & $4^{\rm h} 59.1^{\rm m} $ & $+20^{\circ}
31.9'$ &  31' & 7.7 & & 
& $1.317 \pm 0.270$  & $0.96 \pm 0.14 $ & $2.2{0.45\atop-0.33}$  \\
\cline{3-7} \cline{10-12}
\rule[-1ex]{0pt}{3.5ex} \raisebox{1.5ex}[-1.5ex]{030227} & \raisebox{1.5ex}[-1.5ex]{$\sim$\,20\,s} 
& ISGRI & $4^{\rm h} 57^{\rm m} 32.2^{\rm s}$ & $+20^{\circ} 29' 54"$ & 3'& & \raisebox{1.5ex}[-1.5ex]{21.6'} 
& \raisebox{1.5ex}[-1.5ex]{8.55$^{\circ}$} & $1.1$ & $0.75 $ & $1.85 \pm 0.2$ \\
\hline
\hline
\rule[-1ex]{0pt}{3.5ex}   &   & SPI & \multicolumn{3}{c|}{--} & -- & & 8.6$^{\circ} -$
   & $6.4 \pm 1.0$\,Crab & $137 \pm 8$ $\rm Crab\cdot s$ & -- \\
\cline{3-7} \cline{10-12}
\rule[-1ex]{0pt}{3.5ex} \raisebox{1.5ex}[-1.5ex]{030131} & \raisebox{1.5ex}[-1.5ex]{$\sim$\,150\,s} 
& ISGRI & $13^{\rm h} 28^{\rm m} 21^{\rm s}$ & $+30^{\circ} 40' 33"$ & 2.5' & & \raisebox{1.5ex}[-1.5ex]{0'} 
& 10.6$^{\circ}$ & $\sim 1.9$ (6.5 Crab) & 7 &  \\
\hline
\hline
\rule[-1ex]{0pt}{3.5ex}   &   & SPI & $18^{\rm h} 49.3^{\rm m} $ & $31^{\circ}
46.1'$ & 28' & 8.5& & 
& $10.7\pm1.1$ Crab & $30\pm3$ $\rm Crab\cdot s$ &  -- \\
\cline{3-7} \cline{10-12}
\rule[-1ex]{0pt}{3.5ex} \raisebox{1.5ex}[-1.5ex]{021219} & \raisebox{1.5ex}[-1.5ex]{$\sim$\,6\,s} 
& ISGRI & $18^{\rm h} 50^{\rm m} 25^{\rm s} $ & $+31^{\circ}$ 56' 23'' & 2'&
15.5 & \raisebox{1.5ex}[-1.5ex]{17.6'} 
& \raisebox{1.5ex}[-1.5ex]{9.96$^{\circ}$} & $\sim$\,3.7 & 0.9 & $2.0 \pm 0.1$ \\
\hline
\hline
\rule[-1ex]{0pt}{3.5ex}   &   & SPI & $19^{\rm h} 47^{\rm m} 55^{\rm s}$ &
$+28^{\circ} 23' 49"$ & 13'&  53 & &
&   22.9\,$\pm$\,1.3\,Crab& 386\,$\pm$\,44 Crab\,$\cdot$\,s  & $-$   \\
\cline{3-7} \cline{10-12}
\rule[-1ex]{0pt}{3.5ex} \raisebox{1.5ex}[-1.5ex]{021125} & \raisebox{1.5ex}[-1.5ex]{$\sim$\,24\,s} 
& ISGRI & $19^{\rm h} 47^{\rm m} 56^{\rm s}$ & $+28^{\circ} 23' 28"$ & 2' & & \raisebox{1.5ex}[-1.5ex]{0.5'} 
& \raisebox{1.5ex}[-1.5ex]{7.2$^{\circ}$}  & 14\,$\pm$\,2\,Crab\,(mean)& $\sim
51$ (in 20--& $\sim 2.2$  \\
\cline{3-7} \cline{6-10}\cline{12-12}
&  & PICsIT & $19^{\rm h} 47^{\rm m} 51^{\rm s}$ & $+28^{\circ} 19' 16"$  & 5'& & -- 
& -- & 9\,$\pm$\,1\,Crab\,(mean) &  500\,keV range) & $\sim 3.7$ \\
\hline
\end{tabular}
\end{tiny}
\end{table*}

{\bf GRB021125:} The first GRB  observed within the FoV of INTEGRAL (in this case the
  partially-coded FoV) was the one with the brightest fluence of the six
  GRBs in Tab.\,\ref{tab:SPI-GRBs}  (\cite{grb021125}). 
The GRB occurred during a period of the Cyg X-1 in-orbit 
calibration where SPI was set into a restricted telemetry mode. This was the
  time period allocated for the IBIS-PICsIT photon-by-photon mode calibration, which
  caused high telemetry loads. Only the telemetry, containing the
  technical- and science-housekeeping data and the onboard spectra, with an
aquisition time of $\sim$\,30\,min, were downlinked. Nevertheless it was possible to
determin a GRB--lightcurve and --location by using the counter data of the 19 Ge-detectors 
of SPI's camera, which are sensitive in the broad energy range from $\sim$\,20\,keV to $\sim$\,8\,MeV. 
Due to the high S/N-ratio, with which the GRB was seen in these data, the radius
of the obtained error circle is approximately twice the one of IBIS/PICsIT. The error circles of IBIS \& SPI
are overlaping each other, thus confirming the reported GRB location.
By comparing the count rate of the GRB event with the one of a Crab
  observation, recorded at a similar off-axis angle (6.45$^\circ$), one gets the
  flux and fluence in the 20\,keV\,--\,8\,MeV range in Crab units.
For IBIS it was only possible to extract the mean flux in 
  praticable energy ranges of the instruments (ISGRI: 20--180\,keV, PICsIT:
  180--500\,keV). The mean flux of SPI is $9.2 \pm 1.1$\,Crab and agrees reasonably with the values in 
Tab.\,\ref{tab:SPI-GRBs}. 
The attempt to extract a GRB spectrum from the SPI on-board spectra failed,
  because there is still an uncertainty in the determination of the time
  interval, during which the spectrum was recorded.

{\bf GRB021219:} 
Also the second FoV-GRB  occurred during the payload performance and
verification phase  (\cite{grb021219}). During this time SPI was still in a restricted telemetry
mode, due to the telemetry limitations in the beginning of the mission. In this case
only the multiple detector events and the events which were processed by the
Pulse-Shape-Discriminator (PSD) were downlinked together with the technical- 
and science-housekeeping data. This mode is especially unsuitable for
the analysis of GRBs, because the single events, carrying the energy information below
200 keV are disregarded on board. Thus the same analysis method was used as
described for GRB021125. The GRB location, lightcurve, fluxes and fluence were
determined with the help of the counter data of the 19 Ge-detectors. The
obtained error circle overlaps with the position reported by ISGRI. The flux
and fluence, again in Crab units, agrees with the one determined for ISGRI.

{\bf GRB030131:}  
As the burst occurred during the slew of INTEGRAL, the standard software could
not be used to extract the burst as described in section \ref{grb-analysis}. 
As SPI was in a restricted
telemetry mode, only multiple events and those single events analysed by the
PSD are availabale for the  analysis of this burst.
Due to the weakness of the burst which adds to the problems of analysing
the burst, no reliable position could be derived from the SPI data.

\begin{figure}
   \centering
   \includegraphics[width=0.5\textwidth]{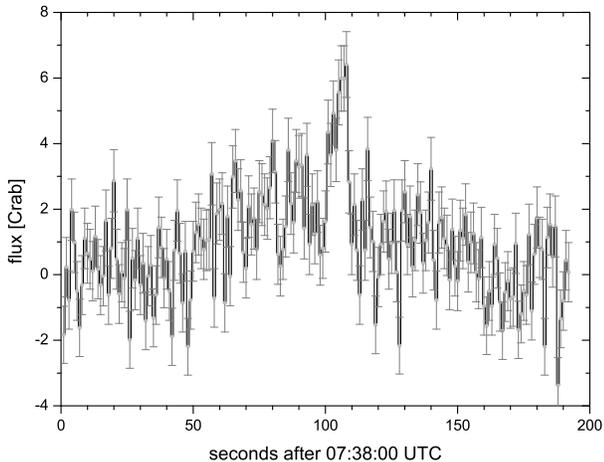} 
   \caption {SPI lightcurve of GRB030131 obtained from the detector count 
rates in the energy range 0.02 -- 8\,MeV, which are part of the scientific-housekeeping data; 
time starts from 07:38:00 UT. The flux in Crab units has been derived by comparing the count rate 
with the one measured for the Crab nebula.}
         \label{fig:grb030131-lc} 
   \end{figure}

But using again the counter data of the 19 Ge-detectors contained in the science-housekeeping 
data, it was possible to determin the lightcurve shown in Fig.\,\ref{fig:grb030131-lc}. 
The burst can bee seen as an increase of the flux starting at $\sim$\,07:38:50 UTC until the 
end of the most prominent peak at $\sim$\,07:39:52 UTC. Probable a part of the next 40\,s  can 
be also attributed to the burst event. Although the spacecraft startet 
slewing at $\sim$\,07:39:06 the derived lightcurve is not affected.   
The peak flux at 10:39:48 UTC agrees well with the one reported by ISGRI (\cite{grb030131}). 
The fluence listed in Tab.\,\ref{tab:SPI-GRBs} was determined for the time period 07:38:50 -- 07:39:52 UTC.

%

{\bf GRB030227:} 
This long (20\,sec) GRB was the first one which was observed in the SPI field of view, while SPI was in 
full operational mode. It was detected by the IBAS program based on the analysis of the IBIS/ISGRI
data (\cite{gcn1895}). Analysis of SPI data on this GRB gave a position $0.36^{\circ}$
off the IBIS position. Even though the burst was weak (total fluence for SPI 
$F = (9.6 \pm 1.4) \cdot 10^{-7} \rm ergs \, cm^{-2}$) a spectral extraction in 2\,s wide bins was 
possible and the SPI data show an indication for hard-to-soft evolution (\cite{Mereghetti2003}). The
overall spectrum has a rather steep slope: $\Gamma = 2.2 {0.45 \atop -0.33}$ ($90 \%$ confidence value).

{\bf GRB030320:} 
So far, this is the GRB observed at the largest off-axis angle, near to edge of
the field of view of SPI (\cite{grb030320-kienlin}). Using the single and multiple detector events SPI was able
to localise the GRB in the 100\,keV to 1\,MeV energy range to an error circle
with a radius of 53'  (99\% confidence value) which agrees with the IBIS position. The
creation of a lightcurve with a 1\,s time binning was only possible when
using the Ge-detector count rates of the science housekeeping data. 
The lightcurve exhibits two prominent peaks during the
$\sim$\,60\,s of prompt emission which are separated by $\sim$\,35\,s.
The flux, fluence and the photon index for SPI was derived by analysing
the count spectrum with an alpha-test version of XSPEC 12.0. It crops up that in
the case of this GRB the XSPEC 12.0 derived results are  in better
agreement, 
compared to the results obtained with the method described
in chapter \ref{grb-analysis}. One has to bear in mind that the instrument
response for sources observed at such large off-axis angles is not fully
understood, because of deficiencies in the knowledge of the mass model.

The hard-to-soft spectral evolution observed for ISGRI is not apparent in the SPI
data, nevertheless  it was possible to confirm the derived ISGRI fluxes of the first and
second peak of emission. 

{\bf GRB030501:} 
This long ($\sim$\,40\,sec) GRB was detected in the partially-coded FoV 
of both, IBIS and SPI (\cite{gcn2183}). 
The position and error circle extracted from the SPI data agree with the one
obtained by ISGRI. The spectral
extraction from both instruments show a similar peak flux, fluence, and spectral shape 
($\Gamma = 1.8 \pm 0.1$ for a single power law). Also comparison with Ulysses (\cite{gcn2187})
and RHESSI (\cite{rhessi}) data showed remarkably consistency. Although the flux seems to be correlated
with the hardness of the GRB spectrum, there is no clear soft-to-hard evolution seen over the duration
of the burst (\cite{beckmann03}).

\subsection{Discussion of the  obtained results}

The first six GRBs observed with INTEGRAL showed that SPI was always able to
detect the same event and to confirm in most cases the results obtained
with IBIS. For the three first bursts it has to be considered that the capabilites
of SPI were weakened by the telemetry limitations in the beginning of the
mission. The next event was the weak GRB030227 and the last two were observed at a
large offset angle, where only 3 to 5 of the 19 Ge-detectors were irradiated by
the GRB. 
So the demonstration of SPI's full capabilities in the case of a strong event in the fully-coded FoV still 
has to take place.

Looking at the current data set  summarised in Tab.\,\ref{tab:SPI-GRBs} most
of the GRBs were detected with a S/N between 7 and 16, with the exception of
GRB021125. At this level SPI is able to localise the GRB down to error
radii of 20' -- 30' and the error circle (90\% confidence) overlapped in most
cases with the one of IBIS. Also the peak flux, fluence and photon indices agreed approximately
with the one derived by IBIS. 

The GRB detection rate in the FoV of SPI of $\sim$\,1 GRB/month in the first six months of the
mission and the obtained positioning errors confirmed the simulations and
calculations carried out before by \cite{skinner97}.

\section{GRB detection with ACS}
\subsection{Detection capabilities}
The ACS provides a large effective area for the detection of bursts (\cite{spie2003}),
but unfortunately with no or only very coarse positional information. The ACS  consists of
91 BGO crystals with a total mass of 512 kg. The energy range for burst detection is
determined by the setting of the energy-discriminators for veto generation, which is
a tradeoff between background reduction and dead-time for the SPI camera. As a veto shield the ACS
has no upper limit of the energy range. 
The sensitivity to $\gamma$-rays depends among other things (e.g projection area) on the attenuation length in the BGO 
crystals which have thicknesses ranging from 16\,mm at the top to 50\,mm at the bottom near to the Ge-camera. 

It is not possible to quote an exact lower energy threshold, due to the
redundancy concept chosen for the ACS hardware. Each of the 90 BGO crystals is viewed by two
photomultipliers (PMTs), which are read out by 90 front-end electronic (FEE) boxes (The $91^{\rm th}$ BGO crystal is
read out by a single PMT and FEE). The redundancy
concept of the ACS is obtained by the cross wiring of neighbouring crystal and electronic boxes. In
this way always anode signals from two PMTs connected to neighbouring crystals are summed in
one FEE. It emerges that a disadvantage of this method is an uncertainty in the energy-threshold
value of individual FEEs, caused by different light yields of neighbouring BGO-crystals and different
PMT properties like quantum efficiency and amplification. A result of this is that the threshold
extends over a wide energy range and is not at all sharp. 
Since the commissioning phase which took place in the first two months after launch the energy discriminators settings 
were set at a constant level which corresponds to approximately $75 {+75 \atop -25}$\,keV.

For the GRB detection on ground with IBAS the ACS housekeeping (HK) data are monitored. These data include
the values of the overall veto counter of the veto control unit (VCU) and the individual ratemeter values of
each FEE. Both kinds of HK data are suitable for burst detection. The count rate of the overall veto counter
(ORed veto signals of all 91 FEEs) is sampled every 50\,ms. A packet, containing 160 consecutive count rates,
is transmitted every 8\,s to ground. If no gap in the telemetry stream occurs one could have a continuous
ACS veto-rate light curve with 50\,ms binning. The measurement time of the individual FEE ratemeter can be
adjusted between 0.1 and 2\,s. An integration time of 1.048\,s has been selected for the mission. 
All 91 FEEs are read out successively in groups of 8 FEEs every
8\,s. The readout of all 91 ratemeter values thus needs 96\,s. In contrast to the VCU overall veto counter
the individual ratemeter values do not yield a continuous stream. Additionally the values of different FEE groups
are shifted by a time interval of 8\,s. So it is very difficult, to derive the burst-arrival direction from
 these individual counting rates.

The SPI/ACS Burst-Alert System  is one branch of IBAS. The trigger algorithm used looks for a significant excess
with respect to a running average, comparable to the trigger algorithm used for other spacecrafts (e.g. Ulysses).
All IBAS processes are multi-threaded applications and run as daemon processes. In the current configuration the ACS
overall veto-counter values are monitored simultaneously at 8 different timescales (0.05\,s, 0.1\,s, 0.2\,s, 0.4\,s, 
0.8\,s, 1\,s, 2\,s and 5\,s). An alert is generated  if two of the monitor programs detect a count-rate excess above
a predefined significance level, which can be set individually for each timescale. The monitoring of the ratemeter
values of individual FEEs is not yet implemented into the ACS part of IBAS but could help to give a rough 
estimation of the GRB arrival direction. This information could
be used to distinguish between the two arrival-cone intersections of the interplanetary network.
The SPI-ACS burst alerts distributed by IBAS are providing the time of occurrence in UTC, the spacecraft
orbital position (in R.A., Dec. and distance to geocenter) and a 105\,s
light-curve (5\,s pre burst trigger, 100\,s after burst trigger) with 50\,ms
resolution. 

Since December 2002 ACS has been added to the $3^{\rm rd}$ interplanetary 
network\footnote{3$^{rd}$ IPN website: http://ssl.berkeley.edu/ipn3/index.html} 
(IPN) of $\gamma$-ray detectors. 
During the first year of the INTEGRAL mission the IPN will consist Ulysses, Mars Odyssey 2001,
Konus-WIND, HETE-II, RHESSI and of course INTEGRAL/SPI-ACS (\cite{hurley97}). The network will have an excellent 
configuration during this time, due to the large spacecraft separations between Earth, Mars and Ulysses, which is orbiting 
around the sun, out of the ecliptic plane.

\subsection{Results obtained for the first bursts observed with SPI/ACS}

We present here the analysis of GRBs detected with the SPI-ACS
during the first eight months (November 2002 - June 2003) of the
INTEGRAL mission. Subsequently, the selection criteria are defined and
the burst sample will shortly be discussed.

\subsubsection{Sample selection}

GRBs show a large variety in appearance, e.g. light-curve
shape and observer-frame duration. In order to allow a decent
statistical analysis, an applicable and robust selection of the sample
of GRB events in the SPI-ACS overall rate is required.

\begin{figure*}[tb]
\centering
\includegraphics[width=5.6cm,angle=-90,bbllx=1.0cm,bblly=1.5cm,bburx=20.0cm,bbury=21.6cm,clip=]{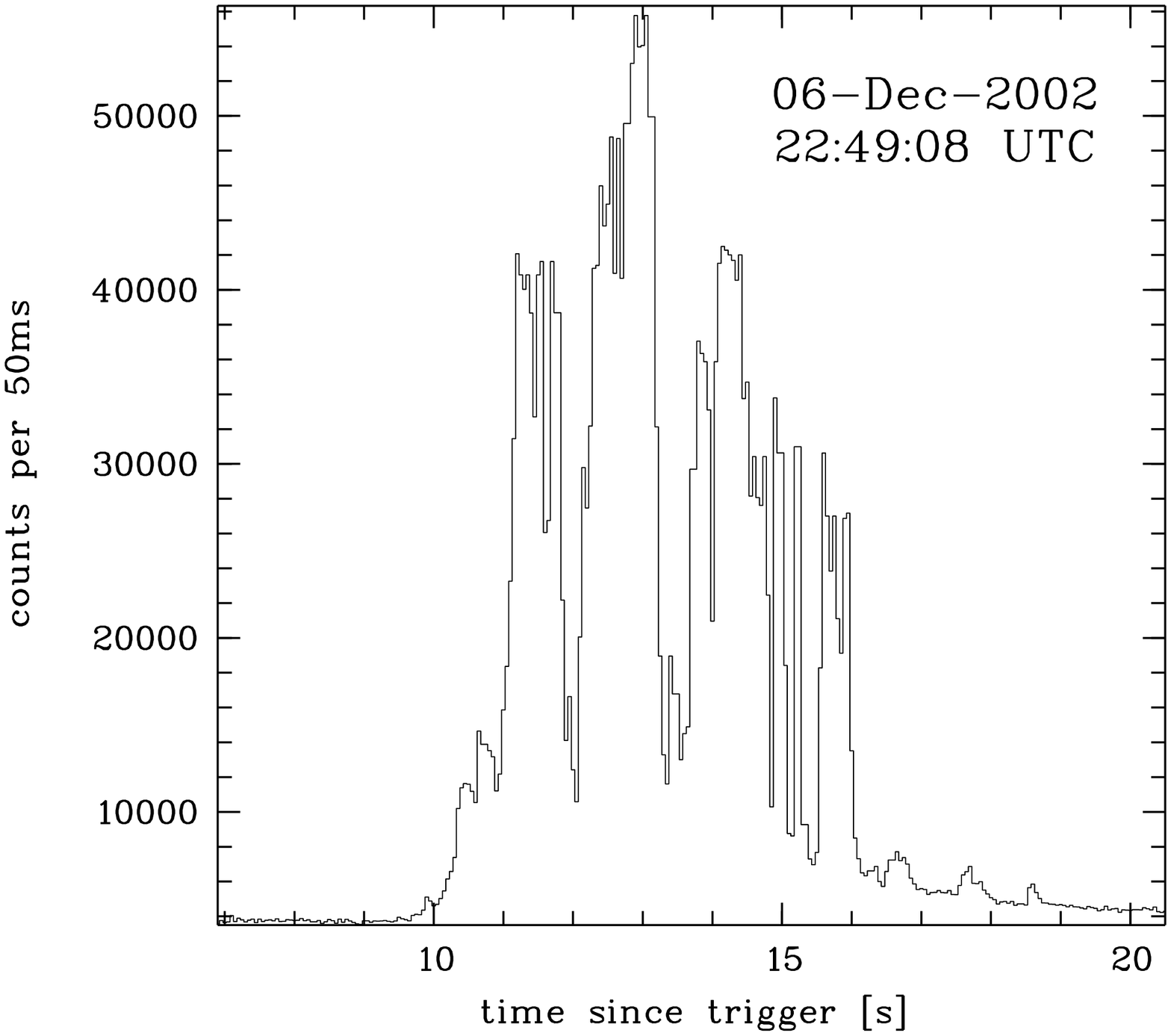}
\includegraphics[width=5.6cm,angle=-90,bbllx=1.0cm,bblly=1.5cm,bburx=20.0cm,bbury=21.6cm,clip=]{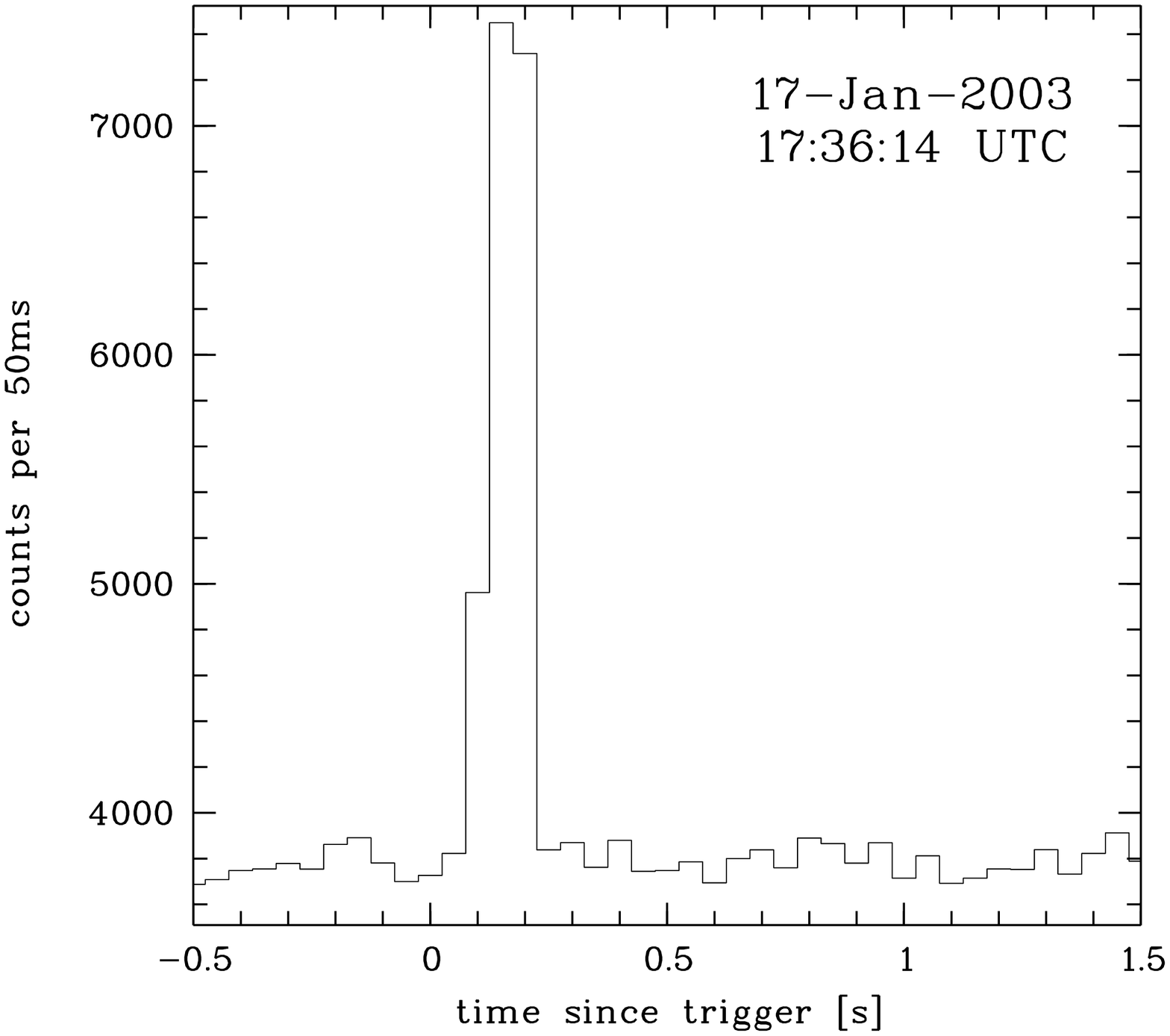}
\includegraphics[width=5.6cm,angle=-90,bbllx=1.0cm,bblly=1.5cm,bburx=20.0cm,bbury=21.6cm,clip=]{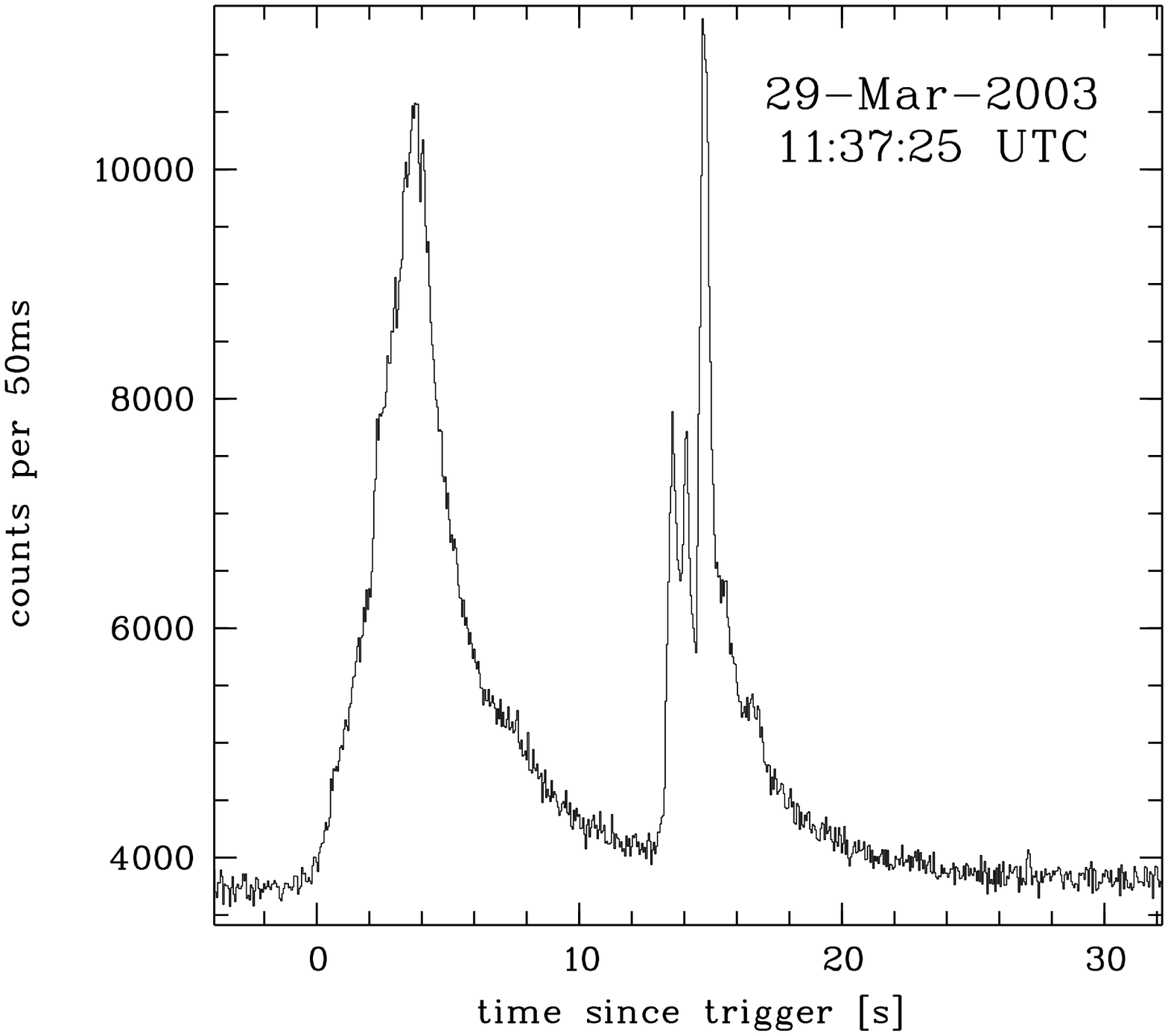}
\caption{Overall rate light curves for three typical GRBs detected by
SPI-ACS with 50\,ms time resolution. We show (from left to right) the
extraordinary bright IPN triangulated burst 021206 (\cite{gcn2281}), 
the short triangulated event from 17th of January
2003 (\cite{gcn1829}) and the extensively studied (\cite{stanek03,hjorth03} and references therein), nearby
(z=0.168; \cite{greiner03}) GRB 030329.}
\label{fig:acsLightcurves}
\end{figure*}

Naturally, the best criterion for selecting GRBs from the variety of events
in SPI-ACS is the localisation on the sky. As SPI-ACS has no spatial
resolution, this selection is fully dependent on the localisation or
confirmation of the respective events by other missions [e.g. IPN, HETE-II]. 
These bursts constitute only a subsample of
the GRBs detected by SPI-ACS as different instrumental properties of the
according missions (e.g. energy range, sensitivity) do not allow the
simultaneous observations/detections of all GRBs of the entire sample. Obviously,
it is worthwhile to study also the sample of bursts which are only visible
in the SPI-ACS rates and not confirmed elsewhere. Probing the very
high energies with the unprecedented sensitivity of the SPI-ACS might open
new insights into the burst populations and burst physics.

We define our sample based on the only property measurable by the SPI-ACS,
the observer-frame light curve. In Fig.\,\ref{fig:acsLightcurves} three
example light curves are shown, representing the known variety of
burst shapes (\cite{Fishman95}).  The selection criteria for our sample are as follows: an
event is categorised as a probable GRB when the total significance above 
the background exceeds a significance level of $\sigma$=12 in at least one
time interval during the event.  E.g., assuming a typical background
value of 3700 counts per 50\,ms and 
taking into account the measured FWHM of the ACS background distribution
which is larger then the expected poissonion distribution by a factor of
1.6, 
e.g. a single 50\,ms event with 4870 total counts
(source+background), a 3x50\,ms event with 4500, 4400 and 4250 counts,
respectively, will match the selection. Each event is subsequently
checked for solar or particle origin using JEM-X and the GOES web
page\footnote{http://www.sec.noaa.gov} and IREM, respectively.
Nevertheless, one has to be aware of the only approximatively known
(due to varying background) biasing against faint short bursts. Thus,
the SPI-ACS GRB sample described here certainly favours the population
of long bright bursts. The complete SPI-ACS GRB sample including
statistics is available 
online\footnote{http://www.mpe.mpg.de/gamma/instruments\\ 
/integral/spi/acs/grb}$^,$\footnote{http://isdc.unige.ch/index.cgi?Science+grb}.

\subsubsection{Analysis and discussion}
According to our sample selection described above, a total of 145 GRB
candidates were detected during the first 8 months of the mission. 58 of these
have been confirmed by other instruments. Using the elapsed mission
time, we find an approximate rate of GRBs detected by the SPI-ACS of $\sim$290 ($\sim$116
confirmed) per year which is in good agreement with the predictions
given in \cite{lichti00} prior to the start of the mission. The total
rate is comparable also to BATSE (\cite{paciesas99}).

In addition to the number of events, the SPI-ACS overall rate
provides the possibility of deriving the burst duration in the
instrumental observer frame and the variability of the light curve. As
no spectral resolution exists, typical burst parameters such as fluence
and peak flux can not be derived. Only the total integrated counts and
the counts in the burst maximum can be extracted from the light
curve. In the following, only the distribution of the instrumental
observer-frame burst duration will be discussed. Other statistical
issues (intensity, variability) will be presented in later publications
and in \cite{ryde2003}.

\begin{figure}
   \centering
   \includegraphics[width=0.32\textwidth,angle=270]{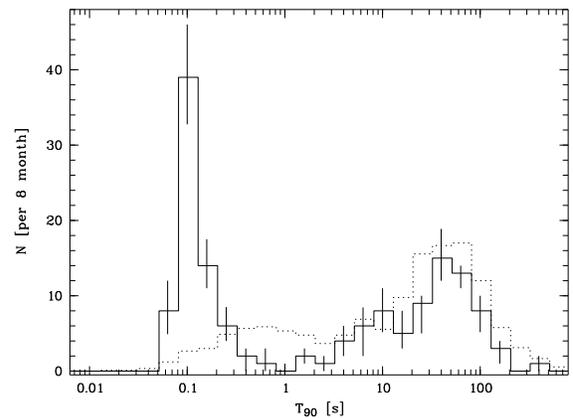} 
   \caption { Distribution of T$_{90}$ for all GRB candidates (solid
line) and for 1234 GRBs from the 4th BATSE GRB catalogue
(\cite{paciesas99}; dotted). In order to compare with the SPI-ACS
detections, the BATSE distribution is scaled to the elapsed INTEGRAL
mission time (8 month). Note the very large fraction of short events 
compared to
BATSE.}
         \label{fig:burstDurationAll} 
   \end{figure}

Fig.~\ref{fig:burstDurationAll} shows the distribution of the measure
for the duration T$_{90}$ (the time interval starting after 5\% and
ending after 95\% of the background subtracted event counts have been
observed) for the sample of SPI-ACS GRBs in comparison to the observed
distribution of 1234 GRBs from the 4$^{\rm th}$ BATSE GRB catalogue
(\cite{paciesas99}). Despite the small sample, a bimodality in the
distribution comparable to that found by BATSE is observed. But
two main differences emerge: i) the SPI-ACS sample contains a
significantly higher fraction of short burst candidates and ii) the
maximum of the short distribution is offset towards shorter duration
for the SPI-ACS sample.

The fraction of short ($<$1\,s) duration GRBs is $\sim$0.48 (70/145)
for the SPI-ACS sample compared to 0.20 for BATSE
(\cite{paciesas99}). As BATSE was observing a softer energy band
(50-320\,keV) and was therefore more sensitive to X-ray rich (long) GRBs
than SPI-ACS, a larger short/long ratio could have been expected for
our sample. What is remarkable is the sharpness of the short
distribution around 0.1\,s. Due to the limited time resolution of
50\,ms the short end cannot be sufficiently defined and resolved by
our data. BATSE was able to trace the distribution to smaller values
of T$_{90}$ due to the better time resolution of 200$\mu$s.

The offset of the maximum for the short events to smaller T$_{90}$
might be due to the different energy bands of SPI-ACS and
BATSE. An apparently shorter duration is measured as it
would be if the bursts would have been observed by BATSE. As T$_{90}$ depends
strongly on the instrumental characteristics and as it is still
unclear how this measure connects to the source frame quantity for a
given burst, the discrepancies are neither surprising nor do they
necessarily  trace different burst populations.
 
  \begin{figure}
   \centering
   \includegraphics[width=0.32\textwidth,angle=270]{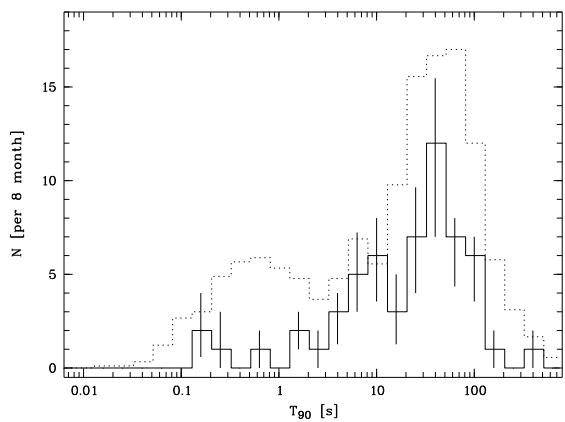} 
   \caption { Same as Fig.~\ref{fig:burstDurationAll} except that
     here only the confirmed SPI-ACS burst sample (solid line) is
     displayed. While most of the long duration bursts are confirmed
     (see Fig.~\ref{fig:burstDurationAll}) for comparison an obvious
     lack of short GRBs can be noticed.}
         \label{fig:burstDurationLoc} 
   \end{figure}

Still, the connection of the short events with real GRBs is not
clear as this population is only marginally observed by other
instruments as shown in Fig.~\ref{fig:burstDurationLoc}. While a large
fraction (73\%; 55/75) of the long bursts are confirmed less than 6\%
(4/70) of the short events were observed by other missions.  This
might be explained at least twofold. On the one hand we might observe
a 'real' short and very hard GRB population, which could so far only
be detected with SPI-ACS due to its high sensitivity at very high
energies. As the current IPN members and HETE-II are generally more
sensitive at lower energies a high fraction of un-confirmed short (and
possible hard) events would not be surprising. These bursts should then
have peak energies above 400\,keV. On the other hand, a significant
contribution to these short events  from instrumental effects
and/or cosmic ray events can not be ruled out. A small contribution
might also arise from soft gamma-ray repeaters (SGRs). SGRs are a small
class of objects that are characterised by brief and intense bursts of
hard X-rays and soft gamma rays. Without localisation SGR bursts can
not be distinguished from short GRBs within SPI-ACS. The issue of
origin of the short events is certainly of high interest and needs a more 
detailed investigation

\section{Conclusions}

SPI confirmed so far all six GRBs detected by IBIS within the FoV.
With the exception of one GRB SPI was able to derive a GRB position, and for GRBs
observed with SPI operating in full telemetry mode, spectra could be extracted.
In one case (GRB030227) some evidence for a hard-to-soft spectral evolution was found
in both, ISGRI and SPI

The ACS of SPI has detected  145 GRBs during the first 8 months. At first preliminary analysis of their duration reveals significant 
differences of their distribution with the one of BATSE. 
The ACS seems to detect very short GRBs (i.e. shorter than 0.2 s).
These bursts are much shorter than the typical short burst, which last for
$\sim 1$ seconds. 
It is unclear if this new ``population'' is real or an artifact of the ACS 
caused by an electrical effect. Also not understood is the deficiency of the normal 
short ($<$\,2\,s) GRBs. It may be due to the fact that the energy threshold of the ACS 
is somewhat higher than the one of BATSE. It is the hope that with the continuing 
observations of GRBs with INTEGRAL these discrepancies can be explained.

\begin{acknowledgements}
The SPI project has been completed under 
the responsibility and leadership of CNES. We are grateful 
to ASI, CEA, CNES, DLR, ESA, INTA, NASA and OSTC for support.
The SPI/ACS project is supported by the German "Ministerium f\"ur Bildung und Forschung" through 
DLR grant 50.OG.9503.0.

\end{acknowledgements}


\bibliographystyle{amsplain}

\begin{thebibliography}{9}
\bibitem[Beckmann 2002]{beckmann02}Beckmann, V. 2002, Proc. XXII Moriond Astrop. Meeting, p. 417, astro-ph/0206506
\bibitem[Beckmann et al. 2003]{beckmann03}Beckmann, V., Borkowski, J., Courvoisier, T.J.-L., et al. 2003, A\&A, this volume
\bibitem[Boella et al.1997]{Boella97} Boella, G., Butler, R.~C., Perola, G.~C. 1997, \aaps, 122, 299 
\bibitem[Coburn \& Boggs 2003]{CobBoggs2003}Coburn, W., \& Boggs, S.E. 2003, Nature  423, 415
\bibitem[Costa et al. 1997]{Costa97}Costa, E., Frontera, F., Heise, J., et al. 1997, Nature 387, 783
\bibitem[Fishman et al. 1989]{Fishman89}
Fishman, G.J., Meegan, C.A., Wilson, R.B., et al. 1989, BATSE: The Burst and Transient Source Experiment on the Gamma Ray Observatory. In Proc. GRO Science Workshop, GSFC, 2
\bibitem[Fishman \& Meegan 1995]{Fishman95}Fishman, G.J. \& Meegan, C.A. 1995, ARAA 33, 415
\bibitem[Galama et al. 1998]{galama98}Galama, T.J., Vreeswijk, P.M., van Paradijs, J., et al. 1998, Nature 395, 670
\bibitem[G\"otz et al. 2003a]{gcn1895} G\"otz, D., Borkowski, J., Mereghetti, S., 2003a, GCN 1895
\bibitem[G\"otz et al. 2003b]{grb030131}G\"otz, D., Mereghetti, S., Hurley, K., et al. 2003b, A\&A, this volume
\bibitem[Granot 2003]{granot03}Granot, J. 2003, ApJL submitted, astro-ph/0306322
\bibitem[Greiner et al. 2003]{greiner03}Greiner, J., Peimbert, M., Estaban, C., et al. 2003, GCN 2020 
\bibitem[Hjorth et al. 2003]{hjorth03}Hjorth, J., Sollerman, J., M{\o}ller, P., et al. 2003, Nature 423, 847
\bibitem[Hurley 1997]{hurley97}Hurley, K. 1997, in Proc.~2$^{nd}$ INTEGRAL Workshop, ESA SP­ 382, 491
\bibitem[Hurley et al. 2003a]{gcn1829}Hurley, K., Mazets, E., Golenetskii, S.,et al. 2003a, GCN 1829
\bibitem[Hurley et al. 2003b]{gcn2187}Hurley, K., von Kienlin, A., Lichti, G. et al., 2003b, GCN 2187
\bibitem[Hurley et al. 2003c]{gcn2281}Hurley, K., Cline, T., Smith, D.M., et al. 2003c, GCN 2281
\bibitem[Kawabata et al. 2003]{kawabata03}Kawabata, K.S., Deng, J., Wang, L., et al. 2003, ApJL accepted,
astro-ph/0306155
\bibitem[Klebesadel et al. 1973]{Klebesadel73}Klebesadel, R., Strong, I., Olsen, R. 1973, ApJ 182, L85
\bibitem[Lichti et al. (2000)]{lichti00}Lichti, G.G., Georgii, R., von
Kienlin, A., et al. 2000, in Proc. 
of the Fifth Compton Symposium, AIP Conf. Proc. 510, 722
\bibitem[Lin et al. 2002]{rhessi}Lin, R.P., Dennis, B.R., Hurford, G.J., et al. 2002, Solar Physics 210, 3
\bibitem[Malaguti et al. 2003]{grb021125}Malaguti, G., Bazzano, A., Beckmann, V., et al. 2003, A\&A, this volume
\bibitem[Mereghetti et al. 2003a]{ibas}Mereghetti, S., G\"otz, D., Borkowski, J., et al. 2003a, A\&A, this volume
\bibitem[Mereghetti et al. 2003b]{gcn2183}Mereghetti, S., G\"otz, D., Borkowski, J., et al. 2003b, GCN 2183
\bibitem[Mereghetti et al. 2003c]{Mereghetti2003}Mereghetti, S., G\"otz, D., Tiengo, A., et al., 2003c, ApJL accepted
\bibitem[Mereghetti et al. 2003d]{grb021219}Mereghetti, S., G\"otz, D., Beckmann, V., et al. 2003d, A\&A, this volume
\bibitem[Metzger et al. 1997]{metzger97}Metzger, M.R., Djogovski, S.G., Kulkarni, S.R., et al. 1997, Nature 387, 879
\bibitem[Paciesas \etal 1999]{paciesas99}Paciesas, W. S., Meegan, C. A., Pendleton, G. N., et al. 1999, ApJS 122, 465
\bibitem[Ryde et al. (2003)]{ryde2003}Ryde, F., Borgonovo, L., Larsson, S., et al. 2003,  A\&A, this volume
\bibitem[Skinner et al. 1997]{skinner97}Skinner, G.K., Connell, P.H., Naya,
J.E., et al. 1997, in Proc. 2$^{nd}$ INTEGRAL Workshop, ESA SP 382, 487
\bibitem[Skinner \& Connell 2003]{spiros}Skinner, G. K., \& Connell, P. H. 2003, A\&A, this volume
\bibitem[Stanek et al. 2003]{stanek03}Stanek, K.Z., Matheson, T., Garnavich P.M., et al. 2003, ApJL submitted, astro-ph/0304173
\bibitem[Ubertini et al. 2003]{ibis}Ubertini, P., Lebrun, F., DiCocco, G., et al. 2003, A\&A, this volume
\bibitem[van Paradijs et al. 1997]{Paradijs97}
van Paradijs, J., Groot, P.J., Galama, T., et al. 1997, Nature 386, 686
\bibitem[Vedrenne et al. 2003]{spi}Vedrenne, G., Roques, J.-P., Sch\"onfelder, V. et al. 2003, A\&A, this volume
\bibitem[von Kienlin et al. 2001]{Kienlin01}von Kienlin, A., Arend, N., \& Lichti, G.G. 2001, 
in Proc. of the International GRB workshop held in Rome, Springer, 427
\bibitem[von Kienlin et al. 2003a]{spie2003}von Kienlin, A., Arend, N., Lichti, G.G., et al. 2003a, in SPIE Conf. Proc. 4851, 
X-ray and Gamma-ray Telescopes and Instruments for Astronomy, 1336
\bibitem[von Kienlin et al. 2003b]{grb030320-kienlin}von Kienlin, A., Beckmann, V., G\"otz, D., et al. 2003b, A\&A, this volume
\bibitem[von Kienlin et al. 2003c]{Kienlin03}von Kienlin, A., Arend, N., \& Lichti, G.G., et al.  2003c, 
in Proc. of the International GRB workshop held in Rome, to be published
\end{thebibliography}


\end{document}